# SIGNAL TRANSCEIVER TRANSIT TIMES AND PROPAGATION DELAY CORRECTIONS FOR RANGING AND GEO-REFERENCING APPLICATIONS

P. KAUFMANN, P. L. KAUFMANN, S. V. D. PAMBOUKIAN, AND R. VILHENA DE MORAIS

**Abstract:** The changes in phase, time and frequency suffered by signals when retransmitted by a remote and inaccessible transponder and the propagation delays are major constraints to obtain accurate ranging measurements in various related applications. We present a new method and system to determine these delays for every single pulsed signal transmission. The process utilizes four ground-based reference stations, synchronized in time and installed at well known geodesic coordinates. The repeater station is located within the fields of view common to the four reference bases, such as in a platform transported by a satellite, balloon, aircraft, etc. Signal transmitted by one of the reference bases is retransmitted by the transponder, received back by the four bases, producing four ranging measurements which are processed to determine uniquely the time delays undergone in every retransmission process. The repeater's positions with respect to each group of three out of four reference bases are given by a system of equations. A minimization function is derived comparing repeater's positions referred to at least two groups of three reference bases. The minimum found by iterative methods provide the signal transit time at the repeater and propagation delays, providing the correct repeater position. The method is applicable to the transponder platform positioning and navigation, time synchronization of remote clocks, and location of targets. The algorithm has been fully demonstrated simulated for practical situation with the transponder carried by an aircraft moving over bases on the ground. The errors of the determinations have been evaluated for uncertainties in clock synchronization, in propagation time delays and other system parameters.

## 1. Introduction

A very well known principle to determine indirectly the distance of a remote object utilizes the echo of a retransmitted signal. When applied to electromagnetic signals the accuracy of this method depends entirely on the knowledge of temporal effects due to four principal causes: (a) the signal speed propagation the medium causing path length variations; (b) propagation time at instruments, cables and connectors at the transmission; (c) propagation time at instruments, cables and connectors at the final reception; and (d) time of signal transit at the remote transponder, which distance is to be determined.

The cause (a) is generally well described by models for propagation in various media (ionosphere, troposphere) as well as in space (see for example [1,6,7,12,15,19], and references therein). The path length variations, however, depends on the elevation angle the object (carrying a repeater) is seen from the bases, which needs to be determined.

Causes (b) and (c) are measured directly, with the high accuracy depending on the quality of the instruments that are utilized. The cause (d) however is undetermined since the remote object, carrying the transponder, is inaccessible for direct measurements. It undergoes changes in its internal signal propagation physical characteristics, which can change with time, or for each sequence of signals used to determine its distance. There are various long distance wireless transmission options using electromagnetic waves, such as the radio waves. The signals are sent to great distances using retransmission repeating links. At these links the signals are received, may or may not be stored or processed, be amplified, and then retransmitted at a frequency which may be the same, or different from the incoming frequency. As mentioned before, the signal transit time at the transponder is affected by several sources.

Other known time changes on fast moving transponders may be neglected for velocities << the speed of light, because they produce effects much smaller in comparison to propagation and delays at the repeater. We refer to Doppler path change caused by frequency shifts in the direction of the repeater [20] and to relativistic effects relative to the reference system containing the sites to which the distances are to be determined, which become further accentuated when the satellites move over distinct gravity potentials relative to the geoid [3,13].

On the other hand, passive transponders, such as the signal scattering reflection by meteor trails in the VHF frequency band [20], may undergo significant phase delays which need to be taken into account for distance measurements.

The precise knowledge of the time changes at each signal interaction at the transponder is an essential requirement for accurate ranging measurements, and applications in remote positioning, navigation, and time synchronization.

## 2. Inference of transit times from the repeater positioning

The knowledge of signal time variations at a repeater is the essential requirement to make feasible the geo-positioning systems and processes using ground based transmitters and receivers and transponders in space. One earlier system and process [8] assumes the use of three reference ground bases and passive scattering of radio signals on meteor trails in the sky. It has been further extended for the use of active signal transponders in space platforms [9]. The concept has been further developed with the introduction of a geometric calculation method [10]. The complete resolving algorithm for those systems, irrespective from any particular geometry, has been given later [14]. The proposed algorithm, however, is critically dependent on the knowledge of signal time changes at the repeater in space, which in practice is not known and on time delays due to propagation in the medium (atmosphere, ionosphere).

To overcome this difficulty a new method and system has been proposed to determine the time variations at a repeater, for every signal interaction [11]. It utilizes four reference bases on the ground (instead of three in the previous systems), A, B, C, and D, as illustrated in Figure 1, where R is the repeater in space, and P a target. All bases, and target P, are synchronized in time, to the best attainable accuracy.

At a given time, a coded signal is transmitted by reference base A, received and retransmitted by R, received back at A, at reference bases B, C, and D, as well as at target P. The ranging measurements obtained for that given instant can be written as:

$$AR(\delta_R, \Delta_{pdAR}) = (\Delta t_A - \delta_{At} - \delta_{Ar} - \delta_R)(c/2) - \Delta_{pdAR} \qquad (1)$$
$$BR(\delta_R, \Delta_{pdAR}, \Delta_{pdBR}) = (\Delta t_B - \delta_{At} - \delta_{Br} - \delta_R)c - AR(\delta_R) - \Delta_{pdBR} - \Delta_{pdAR}$$
$$CR(\delta_R, \Delta_{pdAR}, \Delta_{pdCR}) = (\Delta t_C - \delta_{At} - \delta_{Cr} - \delta_R)c - AR(\delta_R) - \Delta_{pdCR} - \Delta_{pdAR}$$
$$DR(\delta_R, \Delta_{pdAR}, \Delta_{pdDR}) = (\Delta t_D - \delta_{At} - \delta_{Dr} - \delta_R)c - AR(\delta_R) - \Delta_{pdDR} - \Delta_{pdAR}$$
$$PR(\delta_R, \Delta_{pdAR}, \Delta_{pdPR}) = (\Delta t_P - \delta_{At} - \delta_{Pr} - \delta_R)c - AR(\delta_R) - \Delta_{pdPR} - \Delta_{pdAR}$$

to which the respective propagation paths delays $\Delta_{pdAR}$, $\Delta_{pdBR}$, $\Delta_{pdCR}$, $\Delta_{pdDR}$ and $\Delta_{pdPR}$ were added. AR, BR, CR, DR and PR are the distances of bases A, B, C, D and of the target P to the repeater R, respectively, expressed as a function of time variations caused by the signal transit at the repeater $\delta_R$, to be determined, and corrected for the respective propagation path delays $\Delta_{pdAR}$, $\Delta_{pdBR}$, $\Delta_{pdCR}$, $\Delta_{pdDR}$ and $\Delta_{pdPR}$ ; $\Delta t_A$, $\Delta t_B$, $\Delta t_C$, $\Delta t_D$ and $\Delta t_P$ are the time differences effectively measured at bases A, B, C and D, as well as at the target P, respectively, with respect to their clocks; $\delta_{At}$ is the time variation due to signal transit in circuits and cables when transmitted from base A, previously measured and known; $\delta_{Ar}$, $\delta_{Br}$, $\delta_{Cr}$, $\delta_{Dr}$ and $\delta_{Pr}$ are the time variations due to the signal transits on circuits and cables when received at bases A, B, C, D and P, respectively, previously measured and known; and c is the speed in free space of the electromagnetic waves that transport the coded time signal.

The ranging data obtained at the target P are not required for the delay determinations. Since the repeater's position in the sky is not known *a priori*, the time variations $\delta_R$ at the repeater and the path delays can, in practice, be determined simultaneously. As we determine $\delta_R$, the approximate repeater's position is determined, and consequently the elevation angles it is seen from each one of the four bases are defined. The propagation path delays for each distances AR, BR, CR and DR are found utilizing the repeater's elevation angles as seen from the bases applied to a selected propagation model to derive the fully corrected repeater's position. Therefore the values for path delays $\Delta_{pdAR}$, $\Delta_{pdBR}$, $\Delta_{pdCR}$, and $\Delta_{pdDR}$ are added to the system of equations (1), which become a function of $\delta_R$ only. The value of $\Delta_{pdPR}$ will be found similarly but afterwards, when the target position is determined.

The coordinates of the repeater platform R are found in the reference system illustrated in Figure 2, in relation to the x, y and z axis, with measurements obtained at bases A, B, and C, as function of the transit time at the repeater $\delta_R$ [14]:

$$x_R(\delta_R) = \{[AR(\delta_R)]^2 - [BR(\delta_R)]^2 + AB^2\}/[2AB] \quad (2)$$
$$y_R(\delta_R) = \{\{[r_1(\delta_R)]^2 - [r_2(\delta_R)]^2\}/[2y_C]\} + y_C/2$$
$$z_R(\delta_R) = \{[r_1(\delta_R)]^2 - [y_R(\delta_R)]^2\}^{1/2}$$

where

$$[r_1(\delta_R)]^2 = [AR(\delta_R)]^2 - [x_R(\delta_R)]^2 \quad (3)$$
$$[r_2(\delta_R)]^2 = [CR(\delta_R)]^2 - [(x_C - x_R(\delta_R))]^2$$

On the other hand, adding the measurements obtained with bases A, B and D we get:

$$x_R'(\delta_R) = \{[AR(\delta_R)]^2 - [BR(\delta_R)]^2 + AB^2\}/[2AB] \quad \text{(equal to } x_R(\delta_R) \text{ in system (2))} \quad (4)$$
$$y_R'(\delta_R) = \{\{[r_1'(\delta_R)]^2 - [r_2'(\delta_R)]^2\}/[2y_D]\} + y_D/2$$
$$z_R'(\delta_R) = \{[r_1'(\delta_R)]^2 - [y_R'(\delta_R)]^2\}^{1/2}$$

where

$$[r_1'(\delta_R)]^2 = [AR(\delta_R)]^2 - [x_R'(\delta_R)]^2 \text{ (equal to } [r_1(\delta_R)]^2 \text{) in system (3)} \quad (5)$$
$$[r_2'(\delta_R)]^2 = [DR(\delta_R)]^2 - [(x_D - x'_R(\delta_R))]^2$$

The two pair of systems of equations (2) and (3), and (4) and (5) allow us to formulate the discrepancy between the repeater's positions caused by the transit time $\delta_R$ value to be determined given by the following expression [11]:

$$f(\delta_R) = |[x_R(\delta_R), y_R(\delta_R), z_R(\delta_R)] - [x'_R(\delta_R), y'_R(\delta_R), z'_R(\delta_R)]|^2 =$$
$$= [x_R(\delta_R) - x'_R(\delta_R)]^2 + [y_R(\delta_R) - y'_R(\delta_R)]^2 + [z_R(\delta_R) - z'_R(\delta_R)]^2 \qquad (6)$$

The closest value of $\delta_R$ found for every time signal interaction at the repeater corresponds to the minimum value obtained for $f(\delta_R)$ with varying values of $\delta_R$. This can be found by a number of well known numerical calculation methods, such as by iterative procedures or by successive approximations, as it will be in the simulations later in this paper. The accuracy of $\delta_R$ determinations will be in the limit of the instrumental accuracies utilized for the coordinate's determination with the two system of equations (2) and (3), and (4) and (5). The same procedure may be repeated simultaneously, using measurements at bases A, C, and D, improving the accuracy by averaging the two independent estimates for $\delta_R$.

The path delays for each distance AR, BR, CR and DR, caused by propagation in the medium, $\Delta_{pdAR}$, $\Delta_{pdBR}$, $\Delta_{pdCR}$, and $\Delta_{pdDR}$, are derived at the same time calculations using the respective elevation angles the repeater is seen from each one of the four bases, which were determined during calculations.

For the path delay derivation, an adequate propagation model must be adopted, depending on the frequency band used for the coded time signals transmissions. At higher SHF and EHF radio bands, for example, the corrections are due to propagation in the lower atmosphere [6]. The path delay corrections can be approximated using a plane parallel model for the atmosphere

$$\Delta_{pd} = c\, \tau_{atm} / \sin H \qquad (7)$$

where c is the speed of light, $\tau_{atm}$ the atmosphere zenith delay, and H the object elevation angle with respect to the horizon. At low frequency bands, at VHF and UHF and to a certain extent at L-band, the path delay corrections are mostly caused by propagation in the upper atmosphere, ionosphere and plasmasphere [7].

### 3. Applications

#### 3.1. Repeater's navigation

The values of $\delta_R$ and of $\Delta_{pdAR}$, $\Delta_{pdBR}$, $\Delta_{pdCR}$, and $\Delta_{pdDR}$ found with this method are introduced in the system of equations (2) or (4) to obtain the repeater's actual coordinates, for each coded time interaction. The navigation of the platform is immediately derived from successive repeater's positions determinations.

#### 3.2. Synchronism of a remote clock

Every transmitted coded time signal will be also received at other targets, as the example P given in Figure 1. Assuming that the position coordinates of P is known, once the coordinates of the repeater R is determined, the segment PR becomes known in advance. Therefore the discrepancy between the local clock at P and the time received from the clock at the transmitter A can be written as:

$$\Delta t_P = AR/c + PR/c + \delta_{At} + \delta_R + \delta_{Pr} + \Delta_{pdAR} + \Delta_{pdPR} \tag{8}$$

where the segments AR and PR are measured by ranging, $\delta_{At}$ and $\delta_{Pr}$ are experimentally determined and known, and $\delta_R$, $\Delta_{pdAR}$ and $\Delta_{pdPR}$ are determined by the method described here. The clock time correction at P, needed for synchronization will be $\Delta t_p - \Delta_p$, where $\Delta_p$ is the theoretical correct time expected for the clock at position P.

A general illustration for this application is shown in Figure 3. A repeater R is transported by an airplane, which is within the field of view of several actuators labeled L, M, N which geographic positions are known. These may be telecommunications re-transmitters which essential requirement is to operate synchronized to prevent cross-talks and mutual message garbling interferences. The precision of clock corrections can be close to the accuracy attained in the synchronization of reference bases A, B, C and D.

### 3.3. Remote target location

We shall now assume that the remote target P (Figure 2) coordinates are not known, and should be determined by this system and process. The distance from the repeater R, PR, can be determined by a single time interaction, provided that (a) the clock at P is sufficiently well synchronized with respect to the reference bases A, B, C and D and (b) the temporal change of the coded time signal at the repeater $\delta_R$ and the propagation path delays are determined accordingly to the method demonstrated here. The coordinates of P can be determined univocally by obtaining four different measurements of its distance to the repeater R, at four distinct instants, provided that the respective repeaters´ positions are not part of a straight line. The analytical procedures have been given by Kaufmann et al. [14]. The calculations expressed by equations (1), (2) and (3) are repeated referring to one set of three reference bases, including A where the time code transmitter is located (i.e., A,B,C or A,B,D or A,C,D). We obtain the equations for four spheres which intersections gives the coordinates of the target P. Figure 4 shows the situation referred to reference bases A, B and C. The target is located in relation to four distinct positions for the repeater, $R_1$, $R_2$, $R_3$ e and $R_4$. The fourth position might not be needed when the surface of the Earth is taken as the fourth sphere of reference. The calculations are done by adding other reference frame as a function of the repeater's positions in $R_1$, $R_2$, $R_3$ and $R_4$, ($R_1$, u, v, w), where the target coordinates u, v and w are a solution to the system. It has been shown that for three repeater positions we obtain two solutions for the target coordinates [14]:

$$u_P = (PR_1^2 - PR_2^2 + R_1R_2^2)/[2R_1R_2] \tag{9}$$
$$v_p = [(\rho_1^2 - \rho_2^2)/(2\ v_{R3})] + v_{R3}/2$$
$$w_P = \pm (\rho_1^2 - v_p^2)^{1/2}$$
where $\rho_1^2 = PR_1^2 - u_P^2$ e $\rho_2^2 = PR_3^2 - (u_{R3} - u_P)^2$.

The uncertainty for the signal of w is eliminated by adding the fourth sphere (a fourth position for the repeater, or the planet's spherical surface), with radius $PR_4$ (see Figure 4). The position found 4can be converted to the system (A,x,y,z) according to known analytical procedures, to express the coordinates of P in this system, i.e. P ($x_P$, $y_P$, $z_P$). The longitude, latitude and altitude of P are obtained by computing the inverse coordinate conversion $M^{-1}$ ($x_P$, $y_P$, $z_P$). The same results are obtained taking one of the other two reference bases (A, B, D or A, C, D).

## 4. Simulations

### 4.1. The accuracy of the new system and process

A new software has been developed to perform simulations using the algorithm shown here [18]. We used the programming language of MATLAB software, version R2010a [16].

To represent the ground-based reference bases we have selected four cities in São Paulo state, Brazil: São Paulo, Itú, Campinas and Bragança Paulista. The city of Atibaia, in the same state, has been selected to simulate the target P. The known locations for each city are shown in Table 1. The bases, target and estimated location for the repeater are shown in map given in Figure 5.

It has been conceived a transponder carried by an aircraft flying at an altitude of about 6 km close to the city of Jundiaí, also shown in Figure 5. For simulation purpose, we arbitrarily assigned four positions for the repeater, shown in Table 2.

To proceed with simulations the position coordinates shown in Tables 1 and 2 were converted into the ECEF (Earth-Centered, Earth-Fixed) coordinates system, using datum WGS84 [4,5,17] for the reference ellipsoid. The location of a point in this system can be done in coordinates (x,y,z). The origin of axes (0,0,0) is the center of the Earth, the axis Z points to the North, axis X points to Greenwich meridian (where latitude = 0 degrees). The altitude is taken as the distance perpendicular and above the surface of the ellipsoid [4,5,17]. The flow diagrams utilized in the new software applications are shown in Figures 6, and 7.

For the simulations we have adopted a fixed value for the well measured delays due to the signal passing through the electronics, cables, and connectors, both on transmission and on reception at each reference base and at the target. The same value was adopted for all pre-fixed delays, $\delta_{At}$, $\delta_{Ar}$, $\delta_{Ar}$, $\delta_{Br}$, $\delta_{Cr}$, $\delta_{Dr}$, $\delta_{Pr}$ = 0.0001 ms.

In order to compare with the simulated calculations, we have pre-established estimated values for the transmission times from the emitting base to the repeater, from the repeater to the reference bases and the target, for every successive repeater's positions. Adopting a hypothetical delay at the repeater equal to 200 ns ($\delta_R$), c = 299792458 m/s, the total times ($\Delta t_A$, $\Delta t_B$, $\Delta t_C$, $\Delta t_D$ and $\Delta t_P$) can be calculated the following system of equations (derived from equation (1)):

$$\Delta t_A = AR(\delta_R)/c + AR(\delta_R)/c + \delta_{At} + \delta_R + \delta_{Ar} \qquad (10)$$
$$\Delta t_B = AR(\delta_R)/c + BR(\delta_R)/c + \delta_{At} + \delta_R + \delta_{Ar}$$
$$\Delta t_C = AR(\delta_R)/c + CR(\delta_R)/c + \delta_{At} + \delta_R + \delta_{Ar}$$
$$\Delta t_D = AR(\delta_R)/c + DR(\delta_R)/c + \delta_{At} + \delta_R + \delta_{Ar}$$
$$\Delta t_P = AR(\delta_R)/c + PR(\delta_R)/c + \delta_{At} + \delta_R + \delta_{Ar}$$

where the path delays were neglected in this step of the simulations. Adopting the information on the reference bases positions, delays at the bases, and transmission times, it was possible to simulate the whole process and determine the time delay at the repeater and its positions at four successive instants. Once the repeater's positions are determined it was possible to simulate the target position determination.

The first step was to calculate the repeater's position at the initial instant, using bases A, B and C, according to equations (2) and (3), and adopting a given initial value for $\delta_R$. In the next step the calculation was repeated using bases A, B and D, according to equations (4) and (5), and, similarly, using bases A, C and D. We will find three different positions for the repeater ($R_1$, $R_2$ and $R_3$) for each adopted value of $\delta_R$. The error in the repeater's positioning has been calculated by the following equation:

error = $\|R_1-R_2\| + \|R_2-R_3\| + \|R_3-R_1\|$ (11)

This calculation (also known as objective function) has been repeated several times, for different adopted values of $\delta_R$, using the MATLAB fmincon function, which provides the error minimization. The minimization process has been done starting with an initial value for $\delta_R$ and setting an upper and lower range for its variation. For this simulation we have adopted an initial value of 1 μs, with lower and upper limits of 0 and 100 μs, respectively. The correct time delay $\delta_R$ at the repeater is found at the end of the minimization process.

The whole process has been repeated for three more positions of the repeater, to allow the target position determination, according to equation (9). The results obtained were very close to the real pre-established values. The delay obtained at the repeater was of 199.999995 ns, very close to the pre-established value to obtain the transmission times (that was of 200ns). The minimization process gave repeater's position error of about 0.001mm. This first step was a necessary to demonstrate the functionality of the method.

Tables 3 and 4 show the simulated position calculations for the repeater and for the target positions, respectively. Comparing these results with the pre-established values set in Tables 1 and 2 we can see that the approximations are very good. The error in the target altitude, for example, is only 0.002mm. Errors in latitude and longitude are less than $10^{-8}$ arcseconds.

**4.2. Estimate of path delays**

The time delays due to the signal propagation in the medium were introduced in the following step. For the present simulations we shall assume radio frequencies in the several GHz range (for example in the S band), for which the delays are dominated by propagation in the terrestrial troposphere. They can be approximately described by equation (7). We have assumed with Honma, Tamura and Reid [6] a typical sea level $\Delta_{pd}$ of about 2.3 m, for the dry component only. The calculations of delays at the repeater, of the repeater's position, and propagation path delays relative to each base were simulated by simultaneous iterations since there variable are interdependent.

With the initially adopted path delay equal to zero, the successive approximations in the minimization process allowed the determination of $\delta_R$, for which the repeater's position was determined, as well as the elevation angles the repeater was observed from each reference base, and the respective path delay length calculated. The path delay lengths were then added to obtain a new position for the repeater. The process has been repeated until obtaining the convergence of values. Details on this process are shown in the flow diagram of Figure 7. Comparing the values found in this phase with the pre-established values we find that the approximations remain excellent, with discrepancies in the repeater's position and in the target position less than $10^{-3}$ mm (see Tables 3 and 4).

### 4.3. Clock synchronism

The time of arrival of the transmitted time coded signal at a target with known coordinates permits the synchronization of the clock at that location. The time elapsed for the coded time signal to travel from the instant of transmission to the target at known position can be calculated using equation (8). The obtained time difference can be used to synchronize the target's clock to the clock at the transmitting base.

The results obtained from simulations are excellent. Assuming the clocks of the system perfectly synchronized, the ideal error at the receiving location is practically zero (i.e., less than $10^{-15}$ nanoseconds).

### 4.4. Estimate of uncertainties

The practical source of uncertainties of the determinations is related to errors in clocks and/or in delays miscalculations. The two sources can be added together within a certain range of uncertainty.

To perform simulations on the clocks (and/or delays) uncertainties, we first set the determinations of $\delta_R$ and of path delays using the system and method described above. We have next generated random values for clock/delay uncertainties at the four reference bases and at the target, within a plausible time interval. The propagation of errors due to the clock synchronism uncertainties on the repeater's position, on the target location and on the clock synchronism, are exhibited in Figure 8 for ± 5 ns, as an example. The plots correspond to 1000 randomly generated delay values, for four repeater's positions. For each delay it was determined the repeater's position, the target position and the clock synchronization accuracy. The most typical errors found for clock uncertainties of ± 5 ns are of less than 2 meters in the repeater's position, of tens of meters for the target location, and of few ns on clock synchronization

### 5. Concluding remarks

The present geo-referencing system and method of calculations, using four reference bases at known locations on the ground and a remote transponder in space, permits the calculations of time delays of signals transiting by the repeater, and other caused by the signal propagation in the medium. The system and method allows the navigation of the repeater, remote target positioning and remote clock synchronization with very high accuracy. The simulations demonstrate the analytical performance of proposed algorithm on practical application with a transponder carried by an aircraft. The main sources of uncertainties are the clocks synchronism at the reference bases and the target.

A relevant application for this new system is the preliminary determination of satellite orbits. Assuming a repeater placed in an artificial satellite and a favorable geometry its position can be determined for every time mark interaction with the same errors or uncertainties found for repeaters at lower altitudes, shown here. A study on satellite orbital applications is currently being developed for future publication

**Acknowledgements**

This research was partially supported by Brazil agency CNPq.

P. Kaufmann, Universidade Presbiteriana Mackenzie, Escola de Engenharia, CRAAM, São Paulo, SP, Brazil
E-mail address: pierrekau@gmail.com

P. L. Kaufmann, Universidade de São Paulo, Instituto de Matemática e Estatística, São Paulo, SP, Brazil
E-mail address: plkaufmann@gmail.com

S.V.D. Pamboukian, Universidade Presbiteriana Mackenzie, Escola de Engenharia, Laboratório de Geotecnologias e CRAAM, São Paulo, SP, Brazil
E-mail address: sergio.pamboukian@gmail.com

R. Vilhena de Morais, Universidade Federal do Estado de São Paulo, Instituto de Ciência e Tecnologia, São José dos Campos, SP, Brazil.
E-mail address: rodolpho.vilhena@gmail.com

-------------------------------------------------------------------------------------------------------

CAPTIONS TO THE TABLES

Table 1 - Geographic precise locations and altitudes of reference bases and target used in simulations, illustrated in Figure 5.

Table 2 - The repeater's positions and altitudes carried by an aircraft, over the city of Jundiaí (see Figure 5).

Table 3 - Simulated calculations for the repeater's positions, corrected for transit times and path delays added.

Table 4 - Simulated calculations for the target position, corrected for transit times at the repeater and path delays added.

CAPTIONS TO THE FIGURES

Figure 1 - Simplified diagram showing the four reference bases A, B, C, D, the repeater R and the target P.

Figure 2 - Spherical and Euclidean coordinate systems for the reference bases, A, B, C and D, the repeater, R and the target P.

Figure 3 - A time signal repeater R carried by an aircraft referred to reference bases A, B, C and D, can synchronize remote actuators at known geographic positions L. M and N.

Figure 4 - Four repeater positions $R_1$, $R_2$, $R_3$, and $R_4$, and possible target positions, P and P' (after L. Kaufmann e*t al.,* 2006), referred to bases A, B and C. To apply the current method, similar geometries are applicable for bases A, C, D and A, B, D.

Figure 5 - A Google-map showing the pre-selected cities used in simulations, with reference bases A, B, C and D and the target P. The repeater R is carried by an aircraft flying around the indicated city.

Figure 6 – Flow diagrams showing the main routine (a), the target position calculation (b) and clock synchronization (c).

Figure 7 - Flow diagram to calculate the transit time at the repeater, the path delays and the repeater's position at four instants.

Figure 8 - Example of errors caused by clock uncertainties, which may include the path delays errors, for a range of ± 5 ns. In (a) the effect in the repeater position; (b) the error on the target location; and (c) the error in clock synchronization.

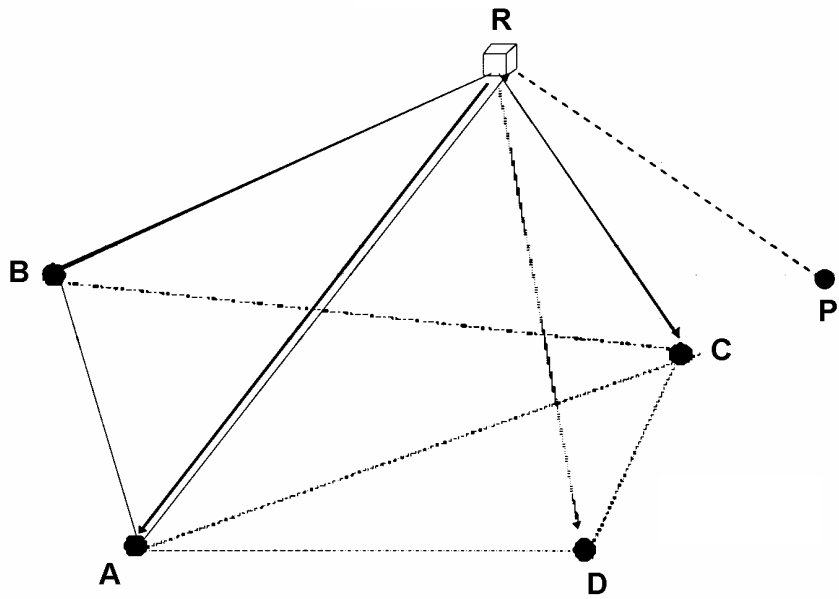

Figure 1

Figure 2

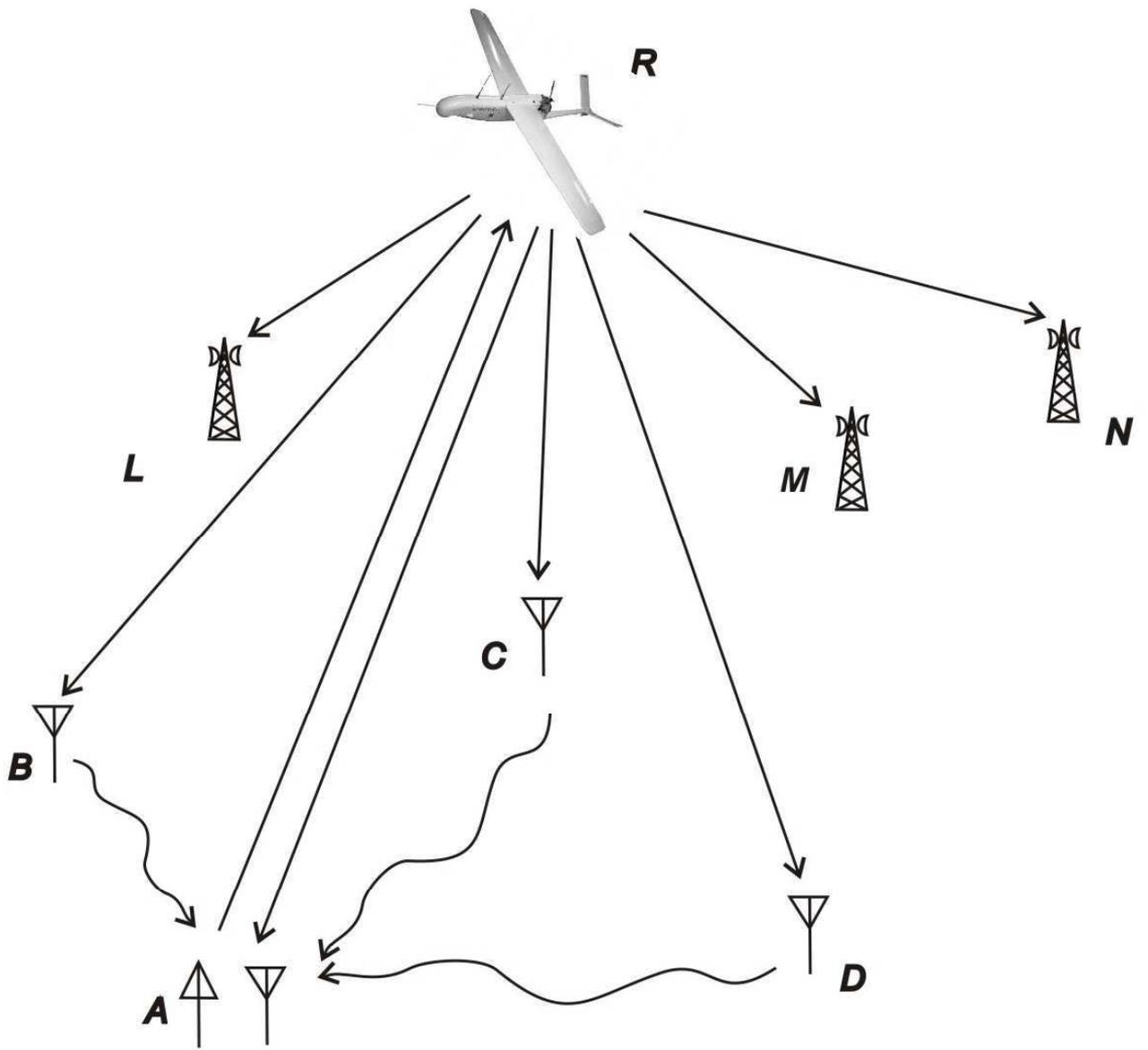

Figure 3

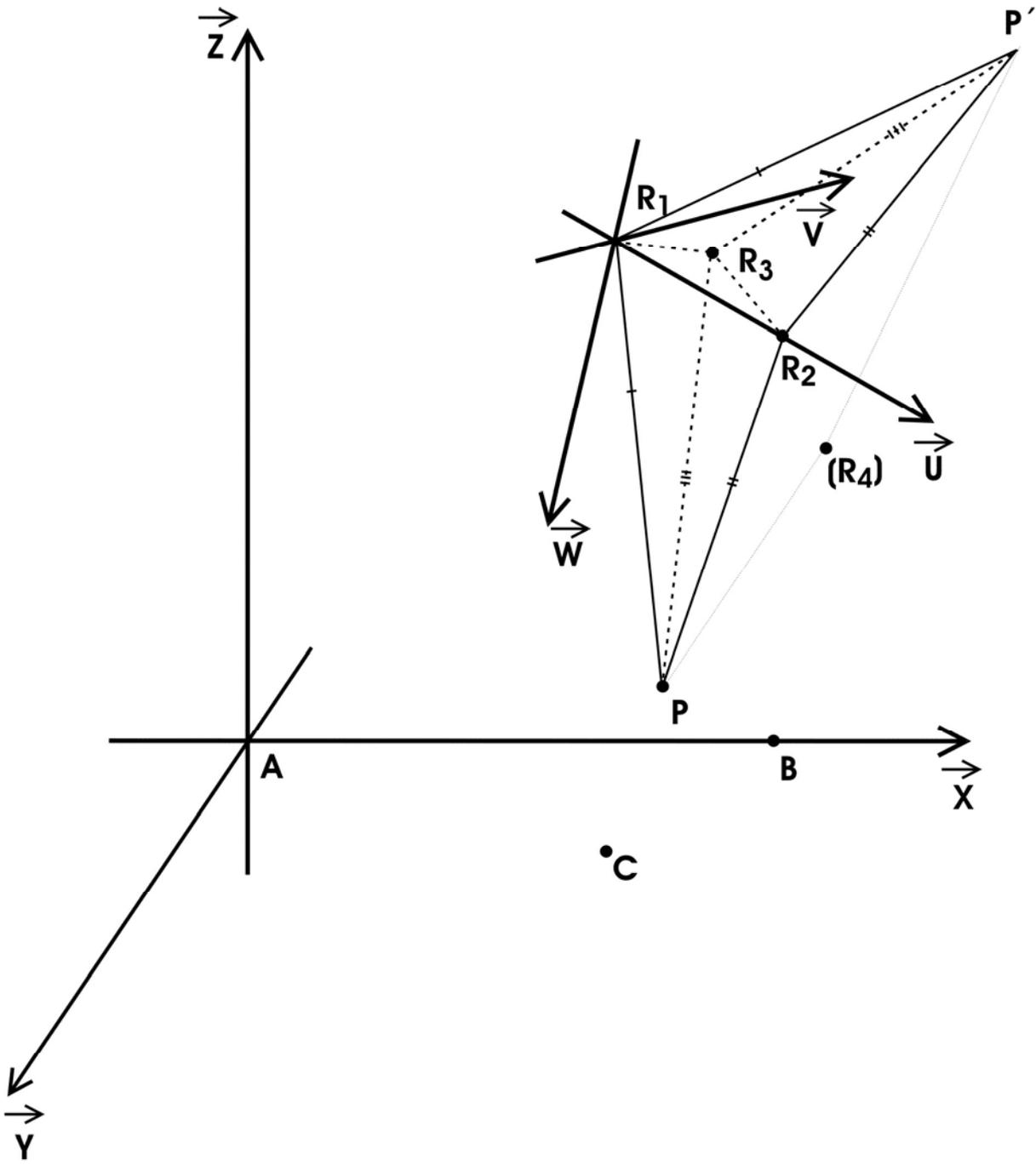

Figure 4

Figure 5

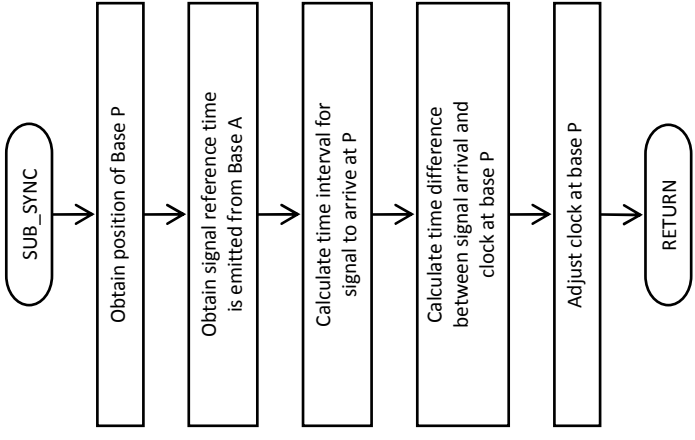
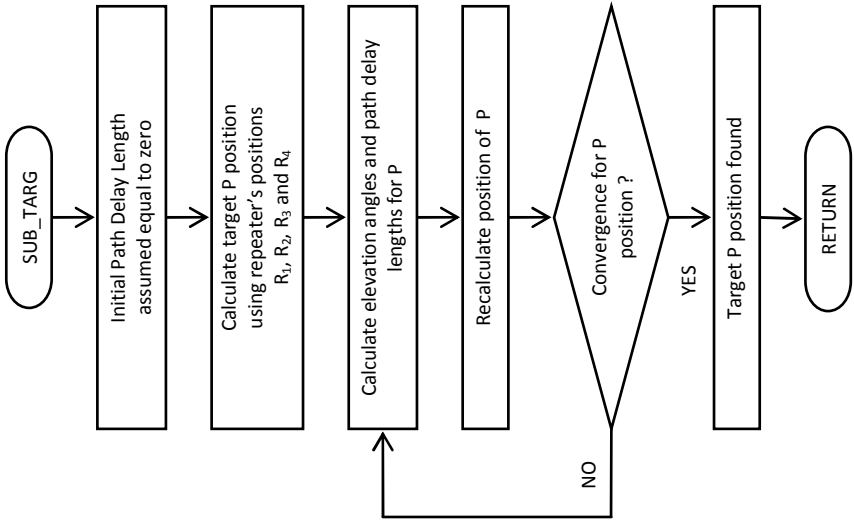
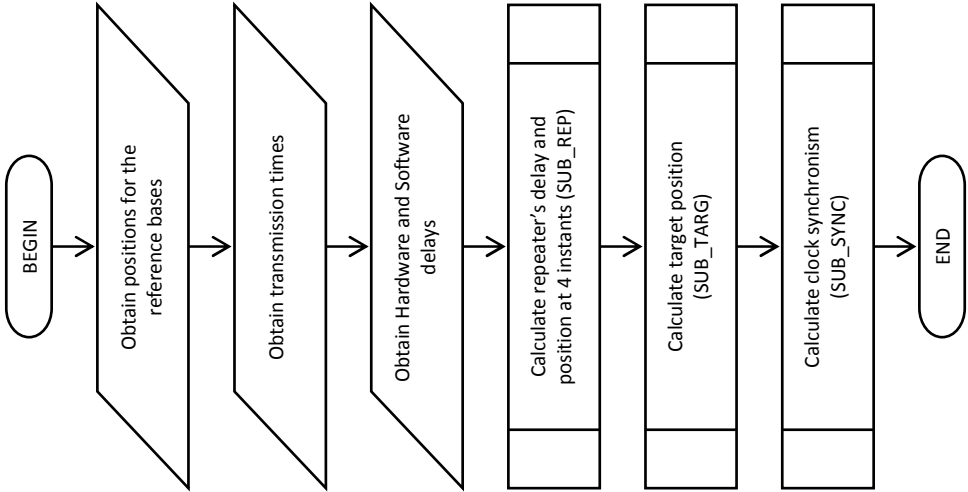

Figure 6

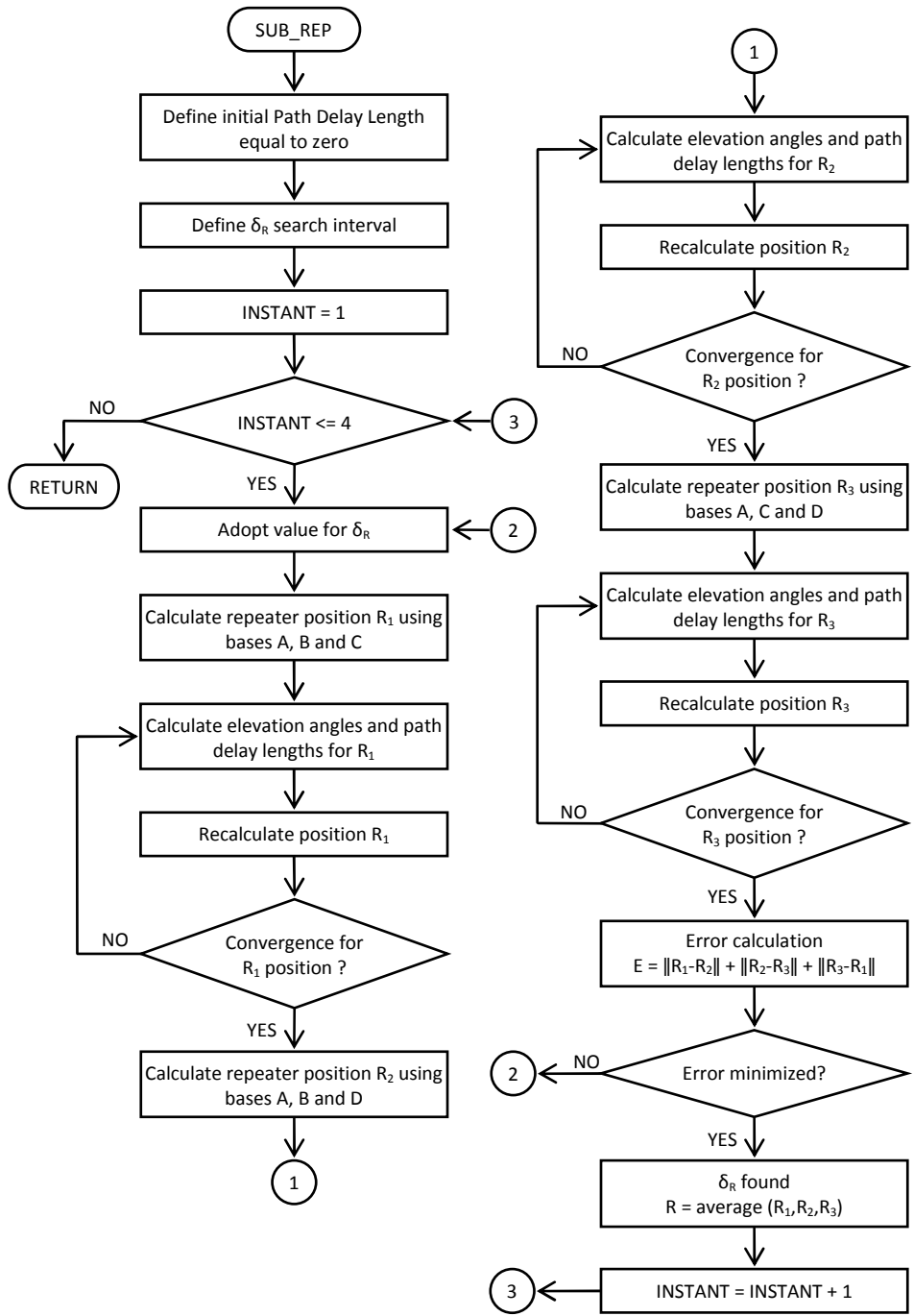

Figure 7

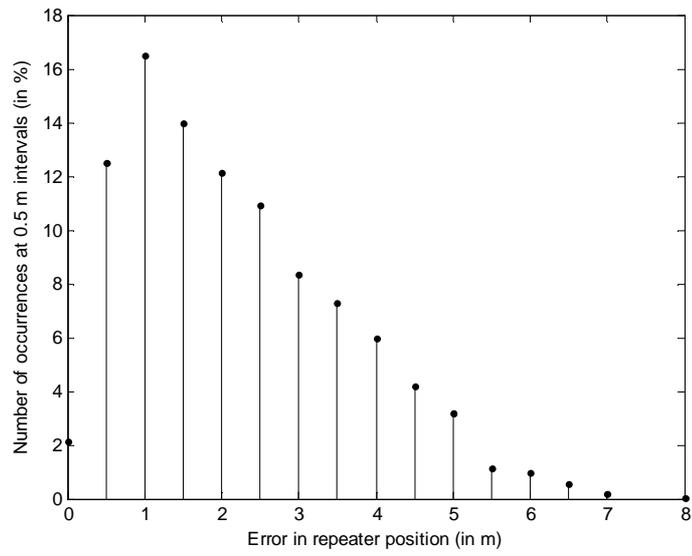

(a)

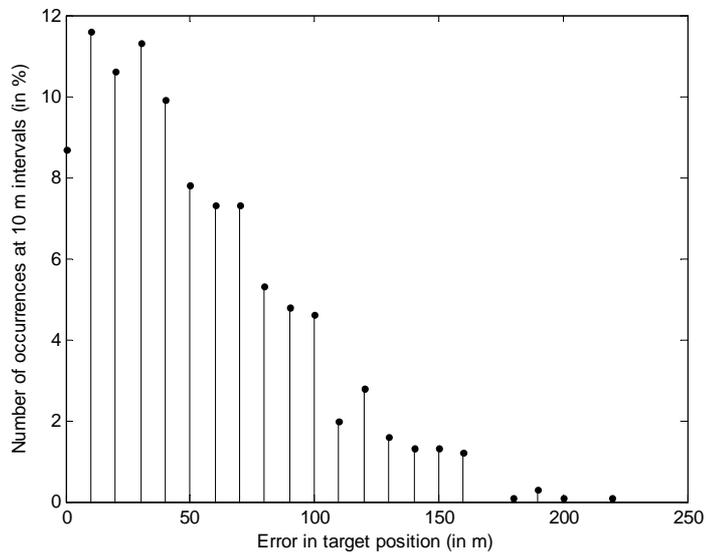

(b)

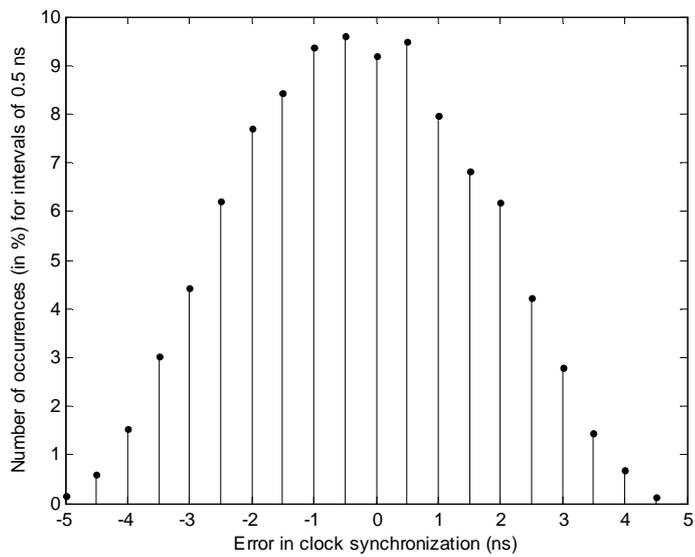

(c)

Figure 8

TABLE 1

| City | Element | Latitude | Longitude | Altitude (m) |
|---|---|---|---|---|
| São Paulo | Base A | -23° 32' 51" | -46° 37' 33" | 730 |
| Itú | Base B | -23° 15' 51" | -47° 17' 57" | 583 |
| Campinas | Base C | -22° 54' 20" | -47° 03' 39" | 855 |
| Bragança Paulista | Base D | -22° 57' 07" | -46° 32' 31" | 817 |
| Atibaia | Target P | -23° 07' 01" | -46° 33' 01" | 803 |

TABLE 2

| City – positions | Latitude | Longitude | Altitude (m) |
|---|---|---|---|
| Jundiaí – Position 1 | -23° 11' 11" | -46° 53' 03" | 5761 |
| Jundiaí – Position 2 | -23° 11' 11" | -46° 59' 03" | 6000 |
| Jundiaí – Position 3 | -23° 17' 11" | -46° 53' 03" | 6200 |
| Jundiaí – Position 4 | -23° 17' 11" | -46° 59' 03" | 6800 |

TABLE 3

| City – positions | Latitude | Longitude | Altitude (m) |
|---|---|---|---|
| Jundiaí – Position 1 | -23° 11' 10.99999999973079" | -46° 53' 03.00000000022578" | 5761.0000008372590 |
| Jundiaí – Position 2 | -23° 11' 10.99999999990985" | -46° 59' 03.00000000035880" | 6000.0000002672896 |
| Jundiaí – Position 3 | -23° 15' 11.00000000065677" | -46° 53' 03.00000000022578" | 6200.0000010607764 |
| Jundiaí – Position 4 | -23° 15' 11.00000000052887" | -46° 59' 03.00000000110060" | 6800.0000007525086 |

TABLE 4

| City | Latitude | Longitude | Altitude (m) |
|---|---|---|---|
| Atibaia | -23° 07' 00.99999999971089" | -46° 33' 00.99999999755028" | 803.00000067241490 |